\documentclass[amsmath,amssymb,aps,prb,twocolumn,groupedaddress]{revtex4-1}

\usepackage{graphicx}% Include figure files
\usepackage{epsfig}
\usepackage{bm}
\begin{document}

%\preprint{}

\title{Universality in the three-dimensional random bond quantum Heisenberg antiferromagnet}


\author{U. Kanbur}
\affiliation{The Graduate School of Natural and Applied Sciences, Dokuz Eyl\"{u}l University, 35160 Izmir, Turkey}
\affiliation{Department of Physics, Karab\"{u}k University, Demir \c{C}elik Campus, 78050 Karab\"{u}k, Turkey}
\author{E. Vatansever}
\email{erol.vatansever@deu.edu.tr}
\author{H. Polat}

\affiliation{Department of Physics, Dokuz Eyl\"{u}l University, TR-35160 Izmir, Turkey}

\date{\today}

\begin{abstract}
The three-dimensional quenched random bond diluted $(J_1-J_2)$ quantum Heisenberg antiferromagnet 
is studied on a simple-cubic lattice. Using extensive stochastic series expansion quantum Monte Carlo 
simulations, we perform very long runs for $L \times L \times L$  lattice up to $L=48$. 
By employing standard finite-size scaling method, the numerical values of the N\'eel temperature
are determined with high precision as a function of the 
coupling ratio $r=J_2/J_1$. Based on the estimated critical exponents, we find that the 
critical behavior of the considered model belongs to the 
pure classical $3D$ $O(3)$ Heisenberg universality class. 
\end{abstract}

%\keywords{}

\maketitle

\section{Introduction}
The concept of disorder and randomness on the critical and universality properties of the magnetic materials 
have great importance for the understanding of the statistical and condensed matter physics  \cite{Young, Sandvik_1, 
Sandvik_2, Wessel_1, Chernyshev_1, Sandvik_3, Sandvik_4, Sachdev_1, Vojta_1, Laflorencie, Lin, Vojta_2, 
Sandvik_5, Sandvik_6, Ibrahim, Liu}. However, clarifying the influence of the disorder and randomness effects on the 
critical exponents of second-order phase transitions has so far been a challenge.  It has been demonstrated 
that there are several approaches to introducing randomness in magnetic materials \cite{Vojta_3}, e.g. the presence of random exchange couplings between interacting spins, or the dilution of magnetic ions. It is a known fact that most of the magnetic materials are more or less defective. 
Therefore, zero- and finite-temperature physical properties of the samples can significantly change depending on 
the kind and amount of defects. They often show unusual and interesting magnetic behaviors that  are  
prominently different from that of their pure counterparts \cite{Vojta_1, Vojta_2, Ibrahim, Sandvik_5}. 
As an example, it has been given in Ref. \cite{Vojta_1} that all critical exponents including dynamical 
correlations are different from the classical percolation values, leading to a novel universality 
class for the percolation quantum phase transition in quantum magnets with quenched disorder.

Among the many spin models, one of the most studied is the Heisenberg model, both from numerical and 
analytical points of view. It builds a strong bridge between the experiments and computer simulations in condensed matter 
physics. In this context, it allows us to explore the underlying physics of the pure and disordered magnetic materials where 
the spin-spin correlations are important. Until the present day, ground state, finite-temperature, and also universality properties of 
the many one-dimensional (1D) and two-dimensional (2D) magnetic systems including different kinds of disorders  
have been comprehensively investigated through a variety of numerical and theoretical methods. Examples include 
quantum spin models with random bonds \cite{Sandvik_1, Sandvik_2, Fisher, Frischmuth, Bergkvist, Hamacher, Yasuda_1, 
Fabritius, Shimokawa}, site dilution  \cite{Wessel_1, Chernyshev_1, Wan, Behre, Kato, Sandvik_7, Yasuda_2, 
Chernyshev_2, Vajk, Yu}, isolated impurity \cite{Eggert}, frustration effects \cite{Albuquerque_1, Bishop_1, Bishop_2, 
Furukawa, Gong, Alet, Stapmanns}, $J-Q$ terms \cite{Sandvik_8, Liu, Iaizzi_1, Iaizzi_2,  Zhao} and 
dimerized $J_{1}-J_{2}$ systems \cite{Albuquerque_2, Wenzel_1, Wenzel_2, Yao, Fritz, Jiang_1, Ran}. 
Magnetic properties of the $S=1/2$ Heisenberg antiferromagnet on an inhomogeneous  2D square lattice 
have been studied by employing quantum Monte Carlo (QMC) simulation in Ref. \cite{Albuquerque_2}.  
The critical exponent $\nu$ of the correlation length is  estimated using finite-size scaling (FSS) analysis, 
which is consistent with the three-dimensional $O(3)$ classical Heisenberg universality 
class $(\nu \approx 0.71)$ \cite{Peczak, Campostrini}. The same confirmation of the $O(3)$ universality class 
has also been shown by Wenzel and Janke in Ref. \cite{Wenzel_2} for the two planar square lattice Heisenberg models
with explicit dimerization or quadrumerization, by making use of the stochastic series expansion (SSE) QMC technique. 
The detailed works in the existing literature  prove that our understanding of critical phenomena of 
the disordered and clean 1D and 2D quantum  magnets has reached a point where well-established results are 
available, as mentioned above. There are, however, quite limited studies on the universality properties of the disordered 
three-dimensional (3D) quantum spin systems. This may be due to the limitation in the 
computational resources that require the averaging of physical observables over a large number of experiments. 

To the best of our knowledge, most of the studies  - inspired by the experimental observation of TlCuCl$_3$  under pressure \cite{Ruegg1, Ruegg2, Merchant}- have been recently performed on different kinds of 3D quantum antiferromagnetic pure and disordered dimerized lattices and have focused on the estimation of the scaling relations of N\'{e}el temperature $(T_{N})$ and the staggered magnetization density ($M_{s}$) near a quantum critical point \cite{Kulik, Oitmaa, Jin, Kao1, Tan1, Tan2, Tan3, Tan4}. Related to this, the first theoretical attempts have been carried out using quantum field theory, and the existence of universal behavior near the quantum critical point has been demonstrated \cite{Kulik, Oitmaa}. In turn, using QMC simulations, Jin and Sandvik have proposed a way to relate $T_{N}$ to $M_{s}$ of the ground state for the several kinds of pure dimerized systems. Analyzing the numerical data, they have found an almost perfect universality \cite{Jin}. In Ref. \cite{Tan1}, universal scaling of $T_{N}$ and $M_{s}$ of the 3D random-exchange quantum antiferromagnets have been investigated within the framework of QMC simulations. The authors have reported that the obtained numerical results support the scaling relations observed for the pure systems for the model including quenched disorder. A similar confirmation of the universal scaling relations  of the relevant physical quantities has been also found for the 3D quantum antiferromagnet with configurational disorder \cite{Tan2}. These studies indicate that the scaling properties of the 3D quantum antiferromagnets are universal in the quantum critical regime, and valid for both pure and disordered models. Since $T_{N}$ and $M_{s}$ are physical observables, the data collapse of these terms has a great significance for experimental physics. Hence, it is evident that most of the attention has been dedicated to clarifying the scaling relations between $T_N$ and $M_s$. There are, however, still unresolved issues regarding the behavior of the 3D quantum Heisenberg model in the existence of quenched disorder. Some of the questions waiting for an answer are given as follows: (i) What is the effect of the disorder on the critical 
temperature of the system? (ii) What will be the universality class of the resulting phase 
transitions? In other words, do the obtained critical exponents depend on the amount of disorder?   In this paper, 
we consider the 3D  random-bond quantum antiferromagnetic 
Heisenberg model with a different perspective.  More specifically, our motivation is to obtain 
an answer for the above questions, and in this way, to determine the universality 
properties of the 3D quantum Heisenberg model in the presence of quenched disorder, employing extensive SSE QMC 
simulations. In a nutshell, our numerical findings indicate that phase transitions of the 3D quantum Heisenberg 
antiferromagnet with quenched disorder belong to the $O(3)$ universality class of the pure 3D classical 
Heisenberg model \cite{Peczak, Campostrini}.

The outline of the paper is as follows:  In Sec. \ref{model2},
we give the model and the details of the simulation scheme.  The numerical results  and discussion are discussed in 
Sec.~\ref{results}, and finally, Sec.~\ref{conclusion} contains a summary of our conclusions. 

\section{Model and Simulation Details}\label{model2}

It is more convenient and compact to express the Hamiltonian of the model in terms of the bond interactions and 
putting restrictions on the created bonds to represent the system of interest. Namely, the Hamiltonian,
\begin{equation}
 \mathcal H=\sum_{b=1}^{N_b}J_b\bm{S_{i(b)}\cdot S_{j(b)}}
     \label{eq:Hamiltonian}
\end{equation}
describes the spin models that include $N_b$ number of bonds with site index $i(b)$ and spin operators $\bm{S_{i(b)}}$ 
interacting with coupling strength $J_b$. For this model, all the interactions are antiferromagnetic ($J_b>0$) and 
among the nearest-neighbor sites on a cubic lattice. These bonds are randomly selected from a bimodal distribution, 
which is given as follows: 
\begin{equation}
\mathcal P(J_{b})=\frac{1}{2}\left [\delta (J_b-J_1)+\delta(J_b-J_2)\right].
\end{equation}

Following Refs. \cite{Malakis1, Malakis2, Malakis3, Vatansever} 
we chose $J_1+J_2=2$ and $J_1>J_2>0$; so $r=J_2/J_1$ defines the disorder strength. It is clear that $r=1$ corresponds 
to the pure 3D spin-1/2 Heisenberg AFM. Among the previously published studies regarding the critical properties 
of the pure model, QMC simulations suggest the location of the transition temperature 
between N\'{e}el and paramagnetic phases as $k_{B}T_{N}/J$=0.946(1) \cite{Sandvik_81} and 0.947 \cite{Troyer}.

\begin{figure}[htbp]
\includegraphics*[width=8.5 cm]{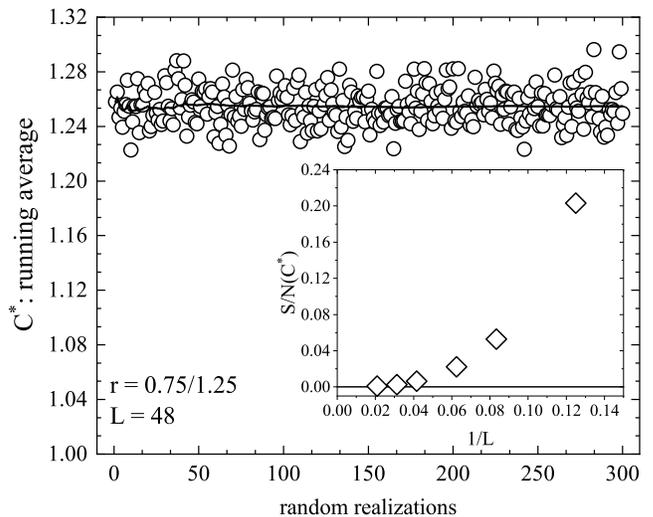}
\caption{\label{Fig1} (Color online) Disorder distributions of the specific heat maxima $(C^*)$ for a 
lattice size $L=48$ for the model with $r=0.75/1.25$. The running averages over the samples are illustrated by a 
solid line. Inset shows the signal-to-noise $S/N$
ratio of the specific heat as a function of the inverse system size. $S/N(C^*)\rightarrow 0$ when
$1/L\rightarrow 0$, indicating that self-averaging is restored in the thermodynamic limit for the 3D bond-diluted 
quantum Heisenberg AFM model \cite{Aharony, Wiseman}.}
\end{figure}

Within the framework of the SSE technique for the isotropic $S=1/2$ Heisenberg antiferromagnetic model, 
the bond operators are split into diagonal and off-diagonal terms as follows,
\begin{subequations}
\begin{equation}
H_{1,b}=\left(\frac14-S^z_{i(b)}S^z_{j(b)}\right) 
\label{eq:diagonal}
\end{equation}
\begin{equation}
H_{2,b}=\frac12\left(S^+_{i(b)}S^-_{j(b)}+S^-_{i(b)}S^+_{j(b)}\right)
\label{eq:offdiagonal}
\end{equation}
\end{subequations}
where $H_{1,b}$ and $H_{2,b}$ are diagonal and off-diagonal bond operators, respectively. 
Then we write the Hamiltonian over the bond operators by indicating random coupling strengths explicitly,
\begin{equation}
\label{eq:bondhamiltonian}
 \mathcal H=-\sum_{b=1}^{N_b}J_b\left(H_{1,b}-H_{2,b}\right)+\mathrm{const.}
\end{equation}
The constant term in Eq.~(\ref{eq:bondhamiltonian}) does not affect the computed quantities  
except for the internal energy and can be included in the post-simulation calculations if desired. 
According to the SSE technique \cite{Sandvik_9, Sandvik_10, Sandvik_11}, 
the partition function is expanded to the Taylor series with a chosen basis,
\begin{equation}
 \label{eq:full_part}
 \mathcal Z=\sum_{\alpha,S_L}(-1)^{n_2}\beta^n\frac{(L-n)!}{L!}\left<\alpha\left|\prod_{p=0}^{L-1}J_{b(p)}H_{a(p),b(p)}\right|\alpha\right>,
\end{equation}
which is a sum over configurations $\alpha$ and all possible operator strings $S_L$ including a unit 
opeartor $H_{0,0}$ to make the length of the strings fixed and a unit bond coupling for a convenient implementation and for this extra index we have $J_0\equiv1$. In this 
scheme, $n$ turns out to be the number of non-unit operators in the string, and $n_2$ is the number 
of off-diagonal operators that appear in pairs in the operator string for a bipartite system thus making 
the matrix elements always positive and $\beta$ is a reduced inverse temperature with Boltzmann constant. 
All the non-zero matrix elements are $J_b/2$ and the weight is given as follows,
\begin{equation}
 \label{eq:weight}
 W(\alpha, S_{L})=\left(\frac{\beta}{2}\right)^{n}\frac{(L-n)!}{L!}\prod_{p=0}^{L-1}J_{b(p)}.
\end{equation}
The relevant quantities are measured for the analysis of thermal and critical behavior by making simulations 
at different system sizes and temperatures for each disorder parameter. In our simulations, $N=L\times L\times L$ defines 
the  total number of spins while $L$ denotes the linear dimension of the lattice, having the values $L=4, 6, 8, 12, 
16, 24, 32$ and $48$. We apply the boundary conditions such that they are periodic in all directions.
The simulations have been realized for three coupling ratios, namely $r=0.9/1.1, 0.75/1.25$ and $0.5/1.5$. For each pair 
of $(L, r)$, we performed 300 independent experiments, generating random seeds. In each sample 
realizations, the first $10^5$ Monte Carlo steps (MCS)\cite{Sandvik_9} are discarded for 
thermalization process, and the numerical data are collected over the next $5 \times 10^5$ MCS. 
After determining the critical temperature region for each 
disorder parameter, the running averages of the specific heat values at a temperature very close to the 
critical point are monitored to ensure the sufficiency of the number of independent realizations. As an 
example, it is clear from Fig.~\ref{Fig1} that $300$ random realizations are found to be enough for the 
statistics of the calculations for the disordered model with parameter $r=0.75/1.25$. We also note 
that the simulations are performed on a cluster with processors Intel(R) Xeon(R) Gold 6148 CPU @ 2.40GHz  
and actual running times in the critical region are measured up to approximately ten days for $L=48$  and a 
single random realization.

Similar averages are also observed for other disorder parameters, which are not shown here. 
The formulation of SSE QMC allows one to derive simple estimators for the quantities of interest. 
The specific heat estimator $(C)$ is defined by the number of 
non-unit operators \cite{Sandvik_10},
\begin{equation}
 \label{eq:specheat}
 C=\left<n^2\right>-\left<n\right>^2-\left<n\right>.
\end{equation}
In the SSE formalism, the Kubo integral can be discretized in a compact form including all the 
propagated states in the imaginary time \cite{Sandvik_11}, that reduces to staggered 
susceptibility $(\mathcal\chi_s)$, which can be defined as follows:
\begin{equation}
 \label{eq:asusc}
 \mathcal\chi_s (\bm{Q})=\left<\frac\beta{n(n+1)}\left[\left(\sum_{k=0}^{n-1} M^z_s(k)
 \right)^2+\sum_{k=0}^{n-1}\left( M^z_s(k)\right)^2
 \right]
 \right>,
\end{equation}
where $\bm Q = (\pi, \pi, \pi)$ is the 3D ordering wave vector while $ M^z_s$ is the staggered 
magnetization, which is given below,
\begin{equation}
M^z_s=\frac{1}{N}\sum_{i=1}^{N}S_{i}^z(-1)^{x_i+y_i+z_i}. 
\end{equation}
The dimensionless Binder parameters are defined as follows:
\begin{equation}
 \label{eq:Q1}
 Q_2=\frac{\left<\left( M_s^z\right)^2\right>}{\left<\left| M_s^z\right|\right>^2},
\end{equation}
\begin{equation}
 \label{eq:Q2}
 Q_4=\frac{\left<\left( M_s^z\right)^4\right>}{\left<\left( M_s^z\right)^2\right>^2},
\end{equation}
and they are independent of the system size at $T_N.$ Another quantity is the spin stiffness that has 
a scaling behavior at the critical point defined as the response to a boundary twist $(\phi)$ and is 
given as (for quantum systems)
\begin{equation}
 \label{eq:stiffness}
\rho=\left.\frac1N\frac{\partial^2E(\phi)}{\partial\phi^2}\right|_{\phi\rightarrow0},
 \end{equation}
where $E(\phi)$ is the energy of the twisted Hamiltonian. The estimator for spin stiffness can be deduced 
from the Kubo integral by averaging a non-diagonal spin current operator \cite{Sandvik_9, Qin}, which gives 
a result in terms of the winding number. For the present model (isotropic) we can improve the value of $\rho$ 
by averaging the estimator for each dimension,
\begin{equation}
 \label{eq:stiffness_estimator}
\rho=\frac{1}{3\beta}\sum_\alpha\langle W_{\alpha}^2\rangle, \; \; \; \alpha \in \{x, y, z\}
 \end{equation}
where 
\begin{equation}
 \label{eq:winding}
\omega_\alpha=\frac1L\left(N_\alpha^+-N_\alpha^-\right),
 \end{equation}
 is the winding number in $\alpha$ direction. The numbers $N_\alpha^+$ and $N_\alpha^-$ count 
 the operators $S^+_{i(b)}S^-_{j(b)}$ and $S^-_{i(b)}S^+_{j(b)}$, respectively, 
 on bond $b$ in the relevant direction. The essence of the SSE technique, 
 implemented for the present and similar systems, is based on the operator 
 loops \cite{Sandvik_9, Sandvik_10} that can take into account winding number 
 sectors exactly thus making the spin stiffness a reliable quantity for 
 critical analysis. 
 
\section{Results and Discussion}\label{results}
\begin{figure}[htbp]
\includegraphics*[width=8.5 cm]{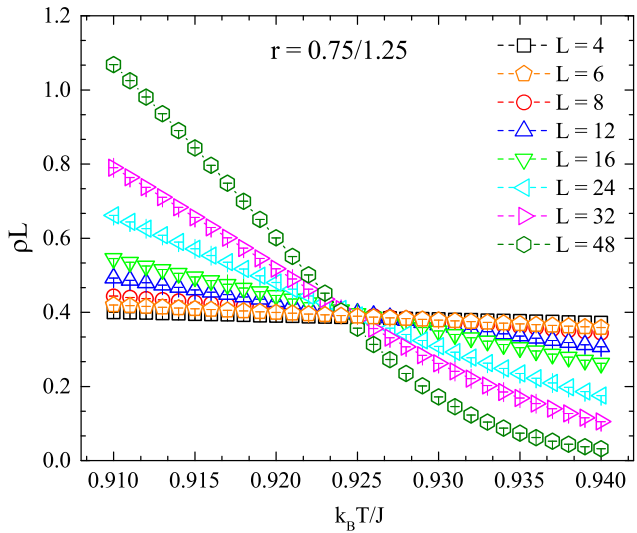}
\caption{\label{Fig2} (Color online) Thermal variation of $\rho L$ for varying values of 
system sizes: $L= 4, 6, 8, 12, 16, 24, 32$ and $48$. 
All curves are obtained for $r=0.75/1.25$. The dashed lines are added to guide the eye.}
\end{figure}

There are several ways to determine the N\'eel temperature $k_BT_N/J$ of the considered system. 
We use three physical quantities: the spin stiffness $(\rho)$, and the 
dimensionless Binder ratios $Q_{2}$ and $Q_{4}$, for all selected coupling ratios $r$. We first consider the
system size and temperature dependence of $\rho$. It should be noted here that $\rho$ can be measured from the global winding number fluctuations in the 
system, within the framework of SSE QMC \cite{Sandvik_12}. According to the hyperscaling theory, finite-size 
scaling of the spin stiffness
can be written as $\rho=L^{2-d-z}$ at the phase transition point \cite{Wallin}. Here $d$ is the dimension of the 
system (as noted before, $d=3$ in our study) while $z$ is the dynamic critical exponent which is zero for finite-temperature phase 
transitions. Therefore, $L\rho$ is independent of $L$ at $k_BT_{N}/J$, which means that the curves versus 
reduced temperature displays an intersection point for two selected different system sizes.  We give thermal variations 
of $\rho L$ for $L=4, 6, 8, 12, 16, 24, 32$ and $48$ at fixed coupling ratio $r=0.75/1.25$, as depicted 
in Fig. \ref{Fig2}. It is clear 
from the figure that $\rho L$ tends to vanish for $k_{B}T/J>k_{B}T_{N}/J$ whereas it begins to increase in the 
range of $k_{B}T/J<k_{B}T_{N}/J$. The curves corresponding to the different pairs of $L$ as a function of the temperature intersects 
almost a single point showing a sign of a phase transition between antiferromagnetic and paramagnetic phases. 

\begin{figure}[htbp]
\includegraphics*[width=8.5 cm]{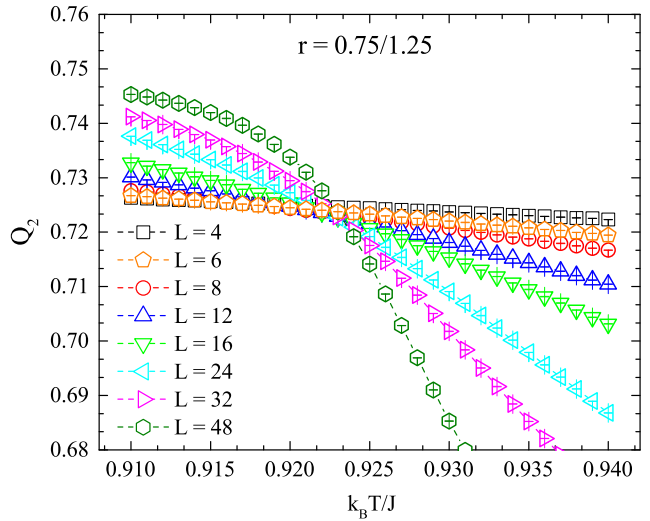}
\caption{\label{Fig3} (Color online) Thermal variation of $Q_{2}$ for varying values of system 
sizes: $L= 4, 6, 8, 12, 16, 24, 32$ and $48$. 
All curves are obtained for $r=0.75/1.25$. The dashed lines are added to guide the eye.}
\end{figure}

\begin{figure}[htbp]
\includegraphics*[width=8.5 cm]{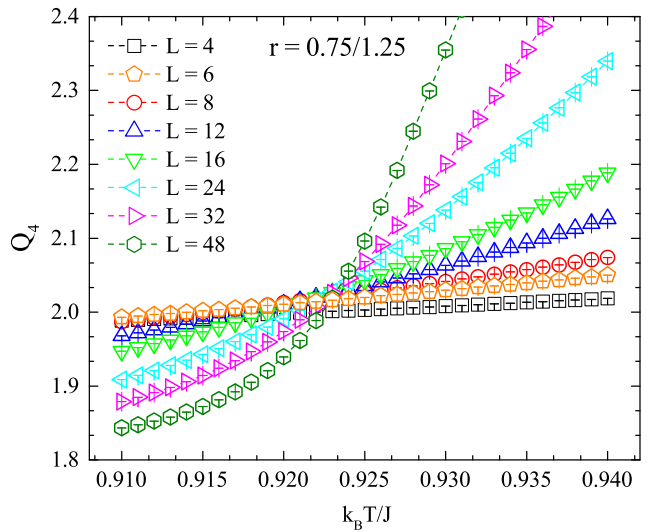}
\caption{\label{Fig4} (Color online) Thermal variation of $Q_{4}$ for varying values of 
system sizes: $L= 4, 6, 8, 12, 16, 24, 32$ and $48$. 
All curves are obtained for $r=0.75/1.25$. The dashed lines are added to guide the eye.}
\end{figure}

In Figs. \ref{Fig3} and \ref{Fig4}, thermal variations of $Q_{2}$ and $Q_{4}$ are depicted 
for the system with the size $L=4, 6, 8, 12, 16, 24, 32$ and $48$ at fixed coupling ratio $r=0.75/1.25$. In these 
figures, we observe that the curves corresponding to varying system sizes tend to cross with 
each other in the vicinity of phase transition of the system.

\begin{figure}[htbp]
\includegraphics*[width=8.5 cm]{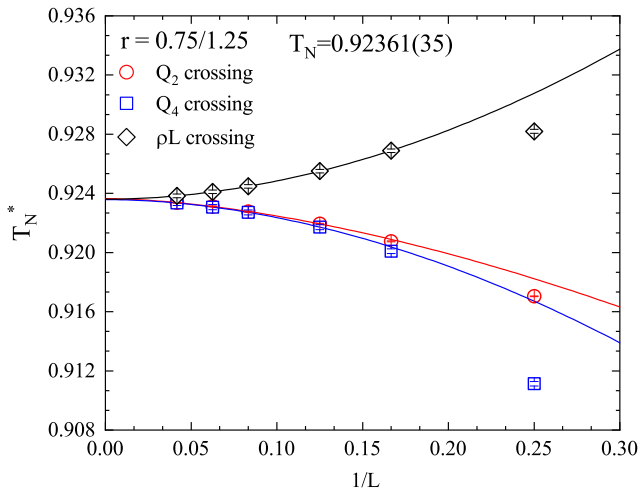}
\caption{\label{Fig5} (Color online) Crossing points of the pair of the system sizes $(L, 2L)$ 
for $\rho L, Q_2$ and $Q_4$  as a function of $1/L$. Lines are fits to the 
equation: $T_{N}^{*}(L)=T(\infty)+aL^{-\omega}$. All curves are obtained for $r=0.75/1.25$.}
\end{figure}

\begin{figure}[htbp]
\includegraphics*[width=8.5 cm]{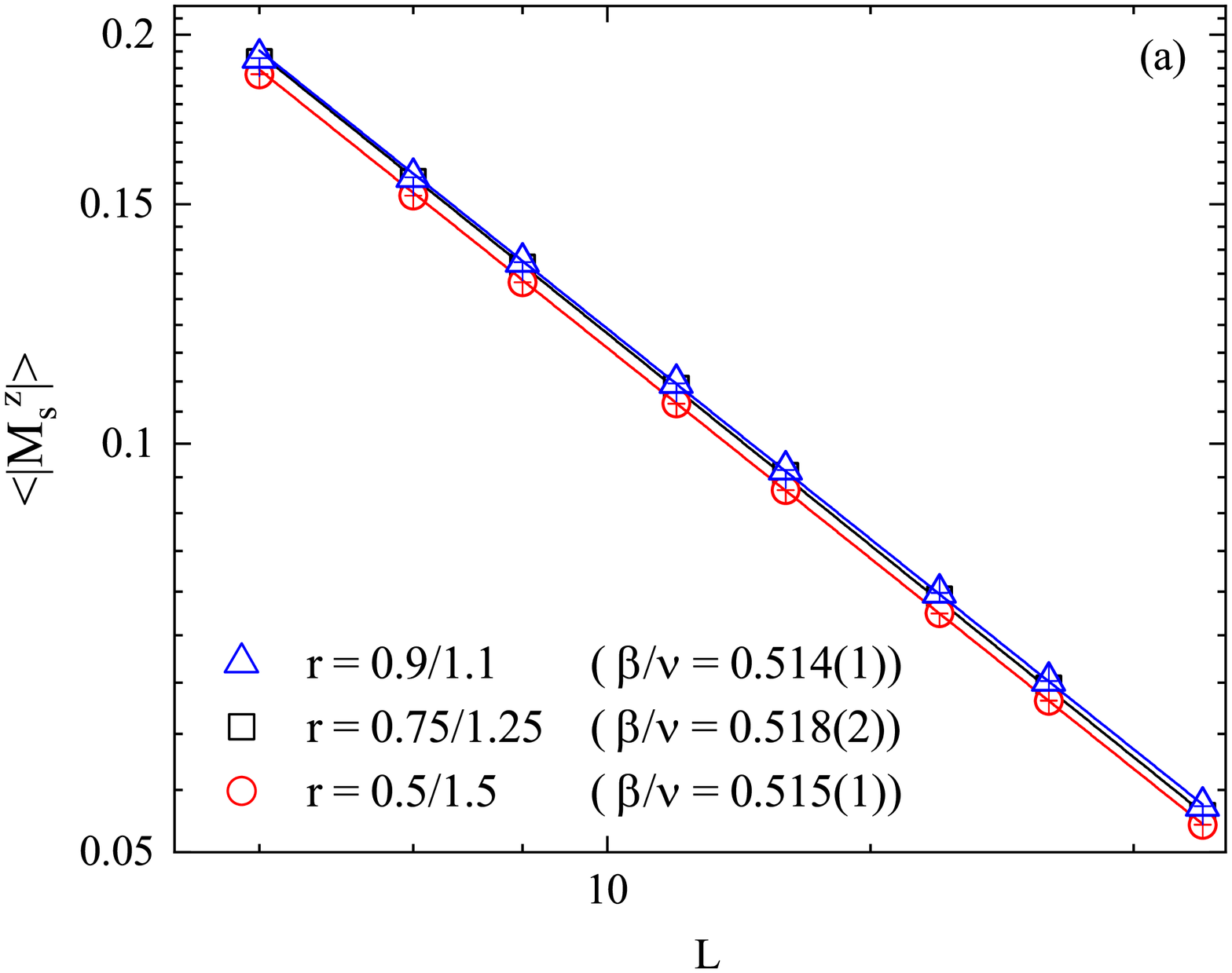}
\includegraphics*[width=8.5 cm]{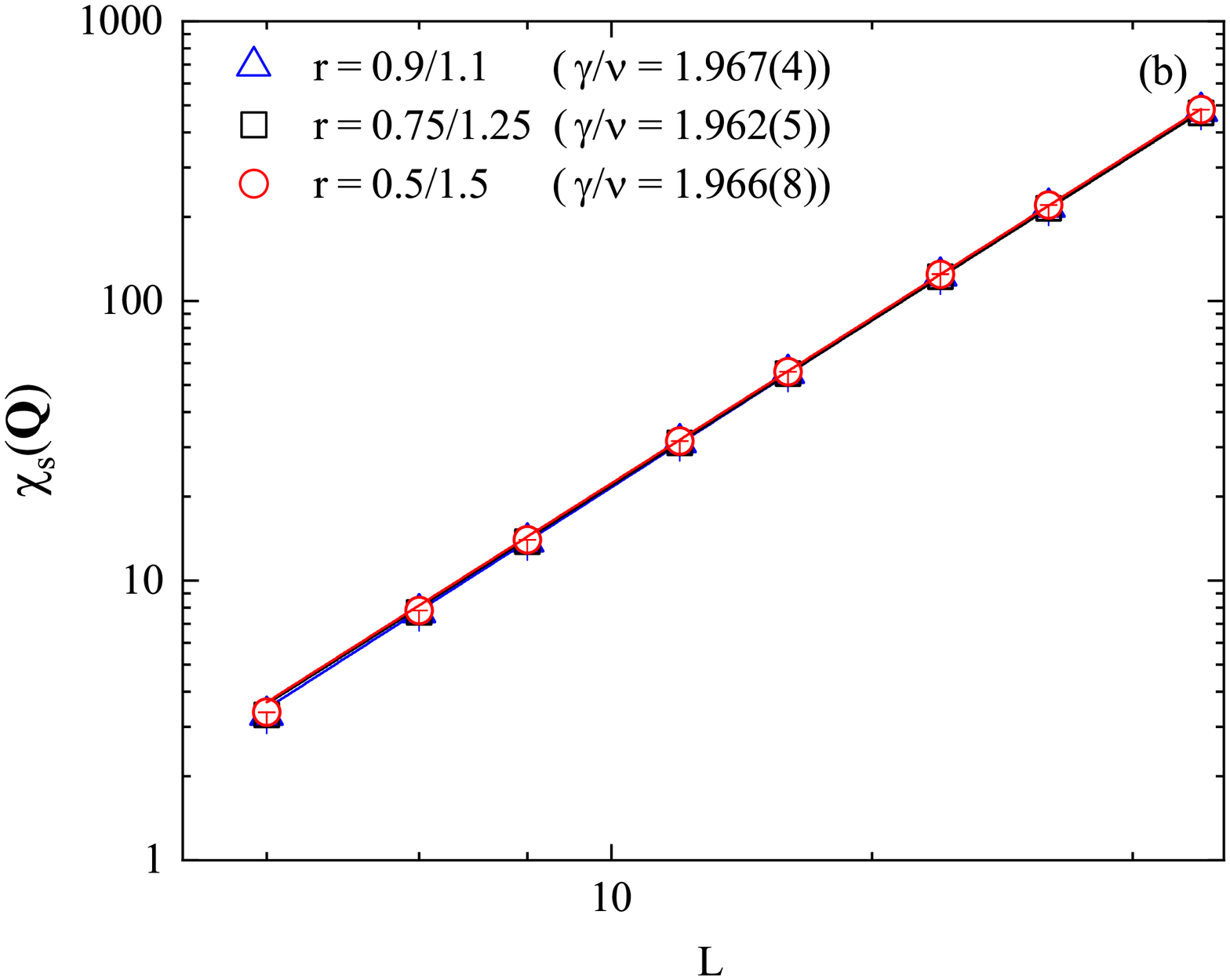}
\includegraphics*[width=8.5 cm]{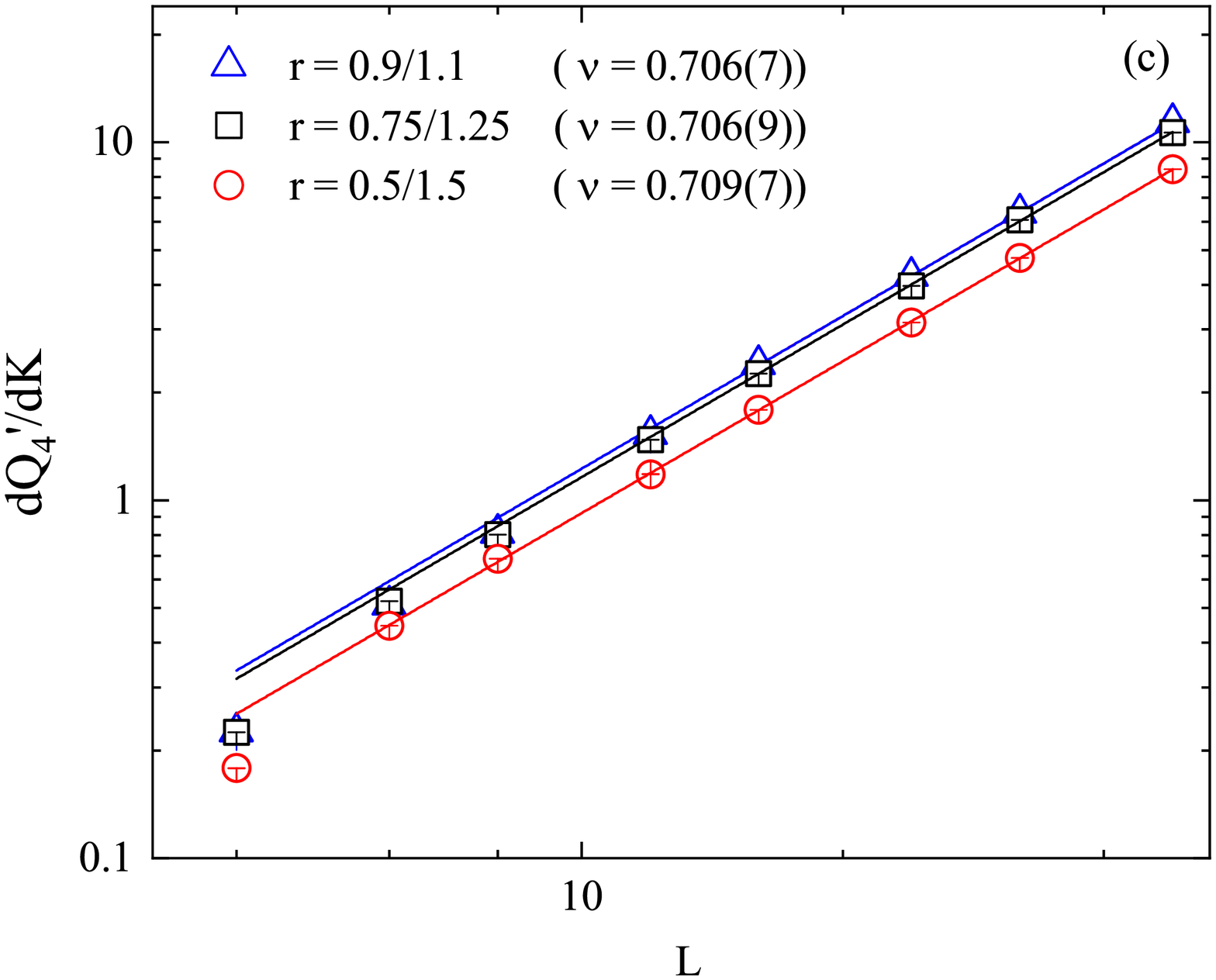}
\caption{\label{Fig6} (Color online) Log-log plots of the magnetization $\langle|M_s^z|\rangle$ (a), staggered susceptibility $\chi_s(\bm{Q})$ (b) and derivative of the Binder cumulant $dQ_4'/dK$ (c) versus the system with the sizes
$L= 4, 6, 8, 12, 16, 24, 32$ and $48$.  All curves are obtained for the considered 
coupling ratios of $r=0.9/1.1$,  $r=0.75/1.25$  and $r=0.5/1.5$. }
\end{figure}

We use the numerical data given for  $\rho L, Q_2$ and $Q_4$ to obtain the intersection 
point of fixed $L$ curves for the pairs of $(L, 2L)$. It is worth noting that 
the numerical data have been collected  using a very small temperature step $\Delta k_{B}T/J=10^{-3}$ for the chosen 
temperature regions. Therefore, we can obtain the intersection point with high accuracy using the 
reliable power-law fits to extrapolate to infinite size, i.e., $L\rightarrow \infty$ limit. We give  $1/L$ dependence of 
$T_{N}^*$, where the relevant quantities cross with each other, as displayed in Fig. \ref{Fig5}. In this figure, 
the solid lines are fits to the function $T_{N}^{*}(L)=T_{\infty}+aL^{-\omega}$. Here, $a$ is a 
constant, and $\omega$ is the crossing point shift exponent \cite{Sandvik_9}. When the relevant data 
points with $L\geq 8$, the scaling function begins to behave well. All the extrapolated values indicate that 
the N\'eel temperature is 0.92361(35) for the fixed coupling ratio $r=0.75/1.25$. Remarkably, it is clear 
from the figure that spin stiffness crossing points converge this value 
from above while the dimensionless Binder ratios converge from below. We note that similar types of observation mentioned 
here have been also found in a variety of quantum spin models. Two notable examples are the  $S=1/2$ Quantum 
Heisenberg bilayers \cite{Wang} and dimerized/quadrumerized Heisenberg models \cite{Wenzel_2}, where the corresponding quantum 
phase transition points have been extracted  in detail, instead of thermal phase transition observed here. 
We obtain the critical temperatures for the remaining selected coupling ratios, 
using the same procedure followed for  $r=0.75/1.25$.  Our Monte Carlo simulations suggest that the critical 
temperatures are 0.94075(3) and 0.861613(12) for $r=0.9/1.1$ and $r=0.5/1.5$, respectively. These results will be used 
to estimate the critical exponents and check the data collapse treatment of the staggered 
susceptibility and scaling behaviors of the system. 

Having determined critical temperatures, we proceed now with testing 
the expected scaling behavior of the system  for all considered coupling 
ratios $r$. According to the standard FSS theory of the equilibrium magnetization $\langle |M_s^z|\rangle$ at 
the critical point, the following power-law can be used to obtain $\beta/\nu$ exponent:
\begin{equation}\label{beta}
 \langle |M_s^z|\rangle\sim L^{-\beta/\nu},
\end{equation}

\noindent where $\beta$ is the critical exponent of the magnetization. 
Figure \ref{Fig6}(a) shows the log-log plots of the staggered magnetization versus the system with 
varying  sizes $L$. Power-law fits of the form of Eq. (\ref{beta}) give the critical exponent ratios
as $\beta/\nu=0.514(1), 0.518(2)$ and $0.515(1)$ for the coupling ratios $r=0.9/1.1, 0.75/1.25$ and $0.5/1.5$, 
respectively. In addition to $\beta/\nu$, we find the critical exponent ratio $\gamma/\nu$  of the 
staggered susceptibility curve, which can be estimated 
by benefiting from the following power-law \cite{Newman, Cardy}:
\begin{equation}
 \chi_s(\bm Q)\sim L ^{\gamma/\nu}.
\end{equation}
\noindent here $\gamma$ denotes the critical exponent of the staggered susceptibility. 
From the log-log plot, our numerical findings suggest that the values of 
the exponent ratios are $\gamma /\nu=1.967(4), 1.962(5)$ and $1.966(8)$ corresponding to the chosen 
coupling ratios $r=0.9/1.1, 0.75/1.25$ and $0.5/1.5$, respectively, as depicted in Figure \ref{Fig6}(b). 
Further evidence can be provided via 
the critical exponent $\nu$ of the correlation length. A simple way to estimate $\nu$ is to use 
the derivative of the Binder cumulant at the critical point. It should obey the relation \cite{Binder}:
\begin{equation}\label{nu}
 \frac{dQ_4'}{dK}\sim L^{1/\nu},
\end{equation}
where $Q_4'=1-Q_{4}/3$ and $K$ is the inverse temperature, i.e., $K=1/k_BT$. Power-law fits of the 
form Eq. (\ref{nu}) are displayed in Figure \ref{Fig6}(c). Our simulation results indicate 
that the critical exponents for the correlation lengths are  estimated as $\nu=0.706(7), 0.706(9)$ and $0.709(7)$
for the studied coupling ratios $r=0.9/1.1, 0.75/1.25$ and $0.5/1.5$, respectively. To briefly summarize, in the view of 
the critical exponents estimated above (i.e., $\beta/\nu, \gamma/\nu$ and $\nu$), it is possible to say that 
the critical exponents  are in excellent agreement with the classical $3D$ $O(3)$ Heisenberg model 
exponents\cite{Peczak, Campostrini}.

\begin{figure}[htbp]
\includegraphics*[width=8.5 cm]{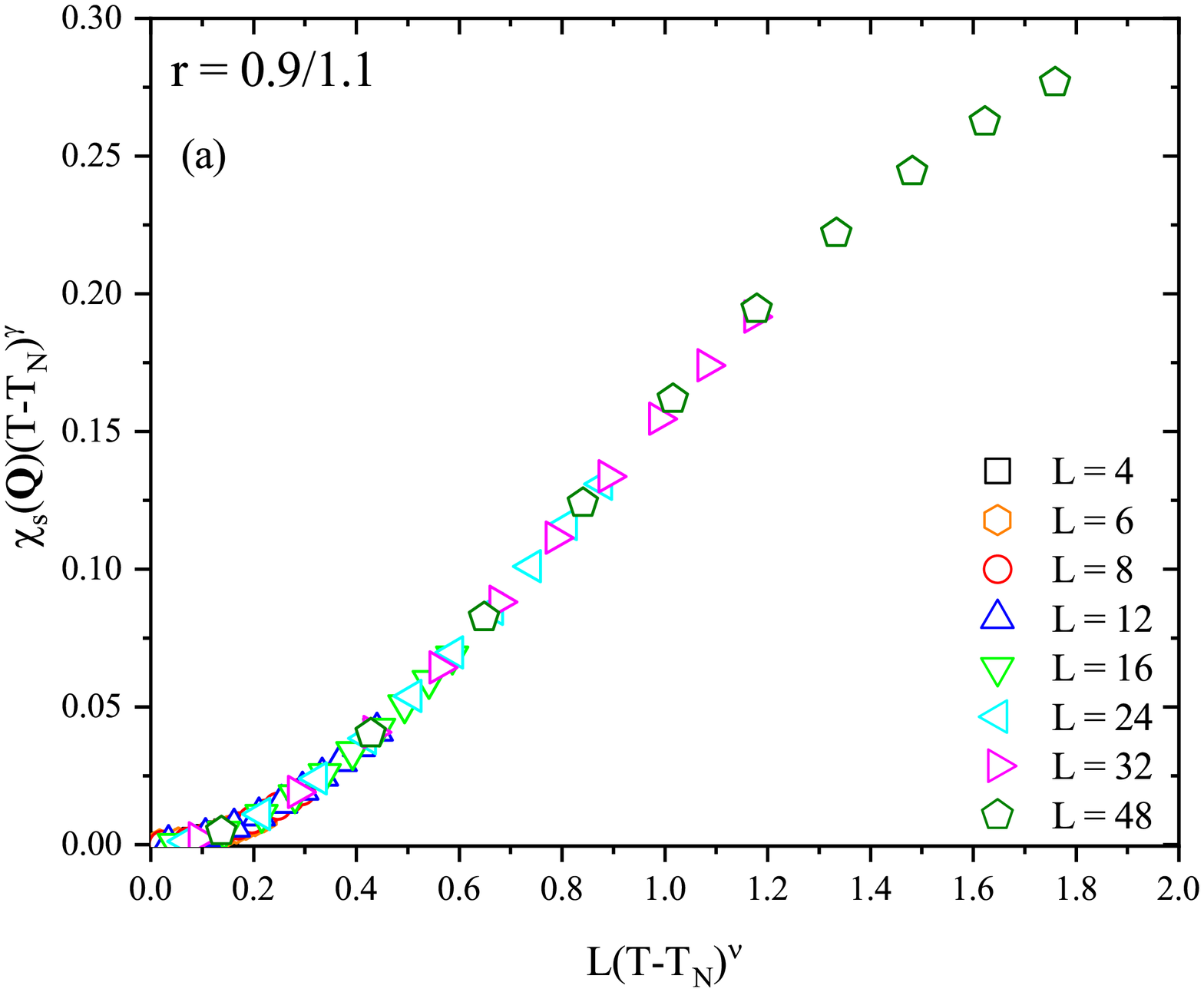}
\includegraphics*[width=8.5 cm]{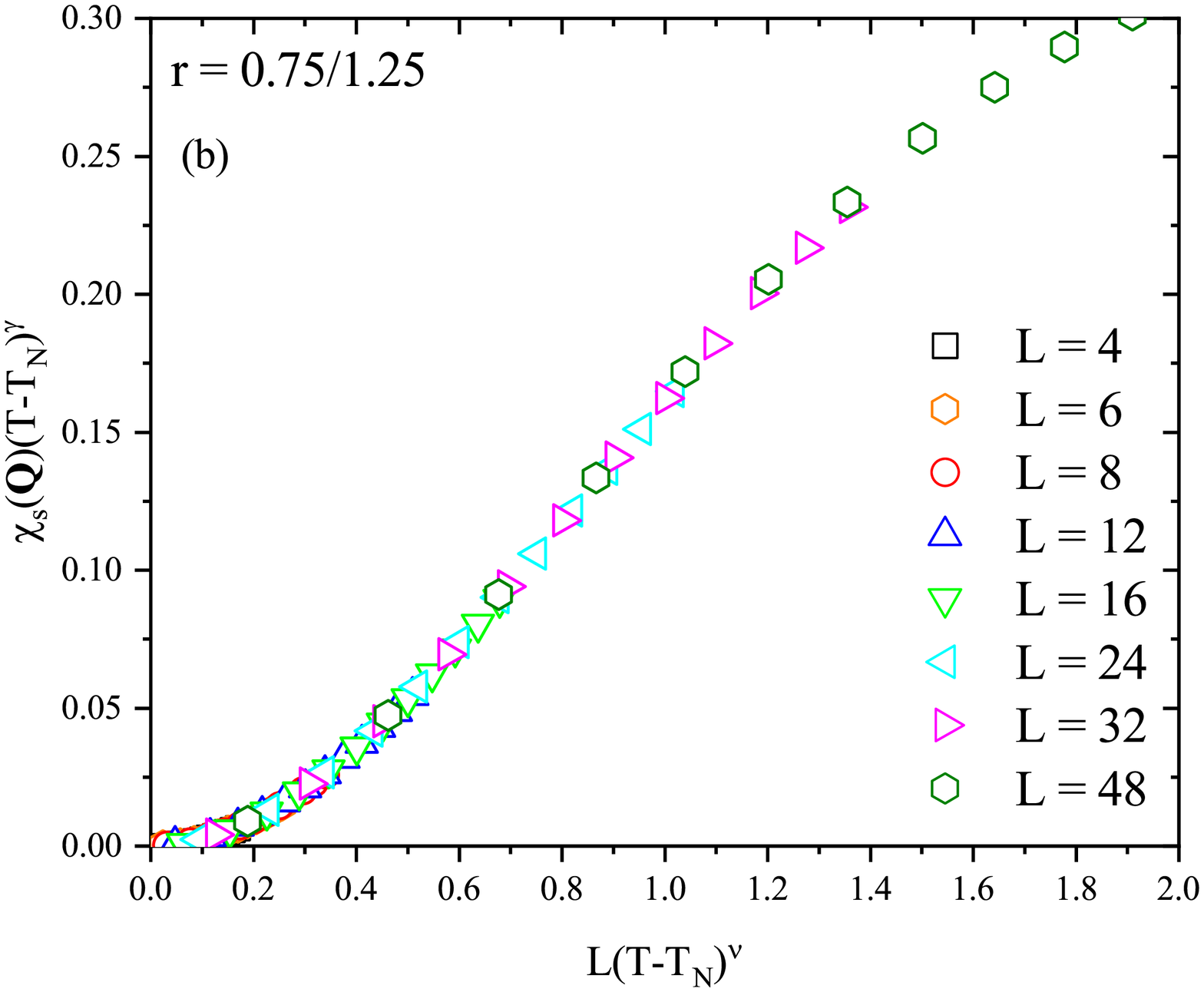}
\includegraphics*[width=8.5 cm]{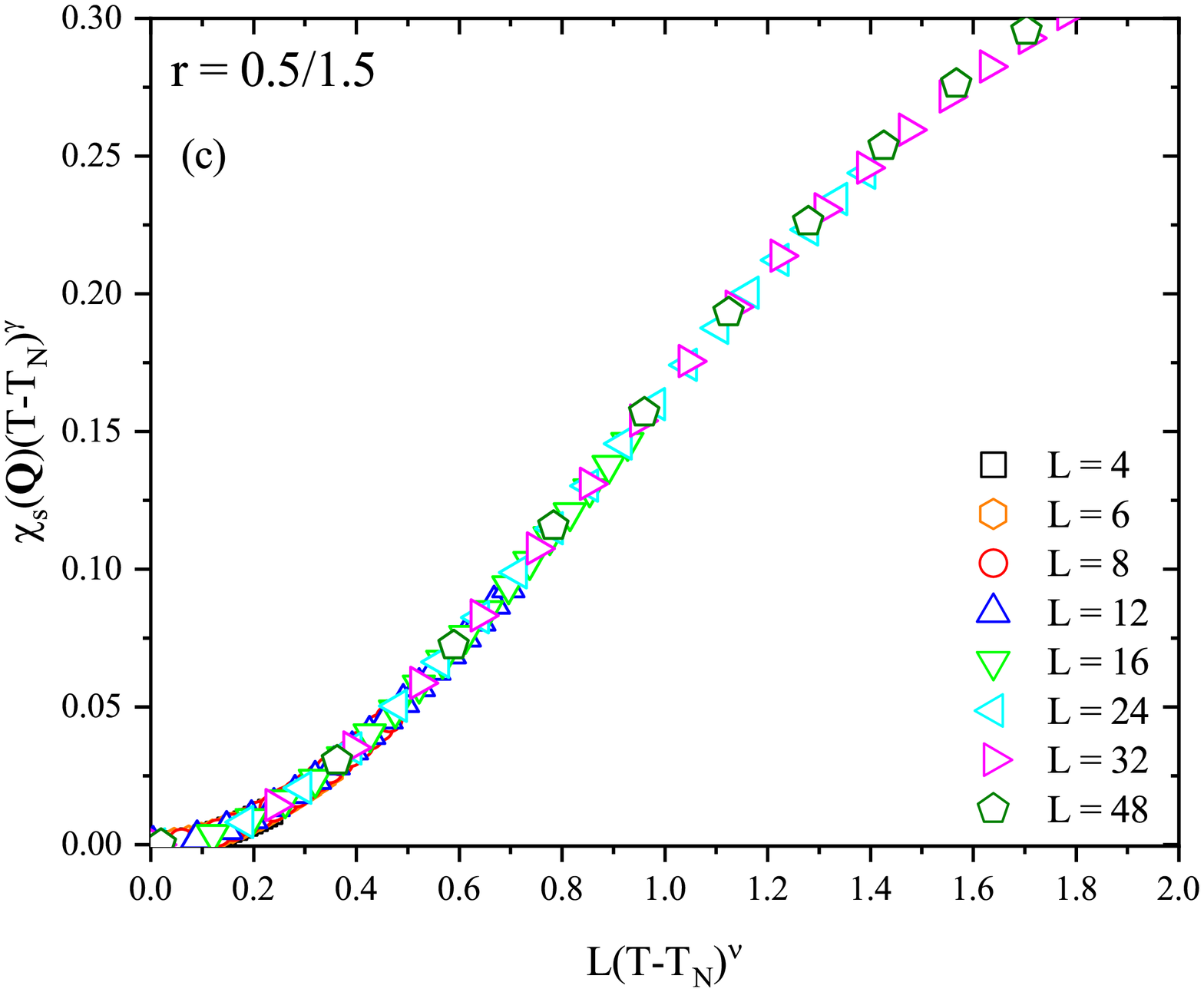}
\caption{\label{Fig7} (Color online) Finite-size scaling of the staggered susceptibility 
above the N\'{e}el temperature for the systems with 
$L= 4, 6, 8, 12, 16, 24, 32$ and $48$. 
All curves are obtained for the considered coupling ratios of (a) $r=0.9/1.1$, (b) $r=0.75/1.25$ 
and (c) $r=0.5/1.5$. The error bars are smaller than the symbols.}
\end{figure}

For a final verification of the $3D$ $O(3)$ Heisenberg universality class, we now continue with testing 
the expected scaling behavior of the system for $T \geq T_{N}$ for all considered coupling 
ratios $r$. In the thermodynamic limit, the susceptibility diverges as $\chi \sim |t|^{-\gamma}$, 
where $t=T-T_{N}$.  Finite-size scaling 
theory also predicts $\chi_{L}(t)=\chi_{\infty}(t)f[\xi(t)/L]$. 
Here, $\xi$ ($\sim t^{-\nu}$) is the correlation length\cite{Newman, Cardy}. 
Based on this definition, it is possible to say that $\chi_{L}(t)t^{\gamma}$ versus $Lt^{\nu}$ curves corresponding to the varying
system sizes should collapse onto a single curve. We give the FSS scaling behavior of the staggered susceptibilities in the vicinity of 
phase transition points of the system with the sizes $L= 4, 6, 8, 12, 16, 24, 32$ and $48$ in Fig. \ref{Fig7}. 
Specifically, we display the curves referring to the varying values of coupling 
ratios $r=0.9/1.1$, $0.75/1.25$ and $0.5/1.5$, in Figs. \ref{Fig7}(a), (b) and (c) respectively. 
For this, $\gamma$ and $\nu$ exponent pairs have been used for each coupling ratio $r$.  Remarkably, 
the data collapse behavior observed here provides a strong indication of the $3D$ $O(3)$ 
Heisenberg universality in the second-order regime of the 3D antiferromagnetic quantum Heisenberg
model in the presence of quenched disorder.

\section{Conclusion}\label{conclusion}
In the present paper we have investigated the effects of quenched disorder on the critical and 
universality properties of the 3D quantum Heisenberg antiferromagnetic model. 
Specifically, we have realized large-scale SSE-QMC simulations on a simple cubic lattice with the system 
size $L$ up to $L=48$ at various values of coupling ratios $r$. First, we have obtained the crossing point of 
$\rho L, Q_2$ and $Q_4$, then present the FSS behavior of the  considered model to estimate the critical 
temperatures with high accuracy for all selected coupling ratios $r$. 
Having determined critical temperatures, we have studied the universality class of the disordered model, 
based on the critical exponents and data collapse analysis. Our simulation results 
indicate that the critical behavior observed for the 
considered model belong to the universality class of the pure classical $3D$ $O(3)$ Heisenberg universality 
class \cite{Peczak, Campostrini}. The results given in this study also strongly support the fact that 
the universality properties of the system do not depend on the microscopic details of 
the Hamiltonian, i.e., spin-spin interactions. 

An important concept concerning the disordered systems is the Harris criterion which states that the 
critical behavior of a pure system is stable against a disorder if the condition $d\nu\geq2$ is satisfied. If the 
inequality does not meet for the pure case of the model, a new universality class with a new correlation 
length exponent $\nu$ satisfying the criterion is expected to emerge \cite{Vojta_4}. The condition $d\nu>2$ holds for the clean case of the present model. Based on the outcomes 
reported in this paper, we conclude that the Harris criterion is not violated since the critical exponents 
found not to change upon the presence of random bond couplings. Hence, 
the Harris criterion provides insight into the criticality of the 3D Heisenberg model with bimodal quenched 
disorder.

Finally, it would be interesting to investigate a new, alternative approach for the 
stacked honeycomb lattices in 3D geometry. A new model considering random 
bond dilution could provide a further understanding of the critical behaviors observed in the 
disordered magnetic systems.

\section*{Acknowledgements}
The authors would like to thank S. Wessel, F.-J. Jiang and D.-X. Yao for many useful comments and discussion on 
the manuscript. The numerical calculations reported in this paper were performed at T\"{U}B\.{I}TAK ULAKBIM, High Performance and 
Grid Computing Center (TR-Grid e-Infrastructure).

%\section*{References}

\end{document}